\documentclass[11pt,a4paper]{article}
\usepackage{authblk}
\usepackage[utf8]{inputenc}
\usepackage[round]{natbib}
\usepackage[pdftex]{graphicx}
\usepackage{amsmath,amsthm,amsfonts}
\usepackage{lscape}   
\usepackage{rotating}
\usepackage{rotfloat}
\usepackage{multirow}
\usepackage{booktabs}
\usepackage{ctable}
\usepackage{threeparttable}
\usepackage{booktabs}
\usepackage{framed}
\usepackage{color}
\usepackage[colorlinks=true,linkcolor=blue,allcolors=blue]{hyperref}

\usepackage{soul}

\title{A note on wavelet shrinkage in nonparametric regression models with ARFIMA errors \thanks{The second author acknowledges financial support of the S\~ao Paulo Research Foundation (FAPESP), grants 2023/02538-0 and 2023/01728-0, and the support of the  the Centre for Applied Research on Econometrics, Finance and Statistics (CAREFS)}}

\author{Alex Rodrigo dos S. Sousa}
\author{Mauricio Zevallos}

\affil{Universidade Estadual de Campinas (UNICAMP)\\ Departament of Statistics, Brazil \thanks{ Sousa (asousa@unicamp.br) and Zevallos (amadeus@unicamp.br)}}

\date{\today} 

\begin{document}

\maketitle

\vspace{-1.0cm}
\begin{abstract}
  In this paper we propose a shrinkage wavelet-based method to estimate the signal in a nonparametric regression model with Autoregressive Fractionally Integrated Moving Average (ARFIMA) errors. Monte Carlo experiments indicate that the proposed method is better than the  universal thresholding rule which is widely used in data analysis via wavelet regression models.\\
  
  \end{abstract}

  

\section{Introduction}
The assumption of independently and identically distributed (iid) Gaussian errors is widely used in nonparametric regression models. In particular, when wavelet-based methods are applied in those models, this assumption in the time domain also ensures iid Gaussian errors in the wavelet domain, i.e., after the application of the discrete wavelet transform to the data. This property offers advantages from an estimation point of view; see \cite{vidakovic-1999}. Although this assumption may hold in practice, and the most commonly used wavelet-based methods are developed under this premise, they are not universal; i.e., random errors may not follow a Gaussian distribution, and, furthermore, may be correlated. This paper addresses a nonparametric regression model under the assumption of long-range (long-memory) dependent errors. Specifically, we assume that the errors follow an Autoregressive Fractionally Integrated Moving Average (ARFIMA) process, as proposed by \cite{granger-joyeux-1980} and \cite{hosking-1981}.

In the literature, sparse attention has been given to nonparametric regression models with correlated errors; see \cite{hall-1990}, \cite{deo-1997}, and \cite{feng-2007} for general approaches involving long-range dependent errors. Furthermore, applications of wavelet-based methods can be found in \cite{wang-1996} who proposed a fractional Gaussian noise model to deal with nonparametric regression with long-range dependency and established asymptotic properties for the minimax risk. In turn, \cite{johnstone-silerman-1997} proposed a level-dependent threshold choice to be applied in the Donoho-Johnstone soft thresholding rule when the data have stationary correlated noise. Recently, \cite{porto-et-al-2016} proposed a wavelet thresholding rule in a model with random design and correlated noise also by adapting the soft thresholding rule. Thus, the above studies considered the thresholding rule in nonparametric regression models under correlated noise, but little attention has been given to Bayesian shrinkage rules in this context. The Bayesian paradigm has the advantage of allowing the incorporation of prior information about the wavelet coefficients in the estimation process, such as the sparsity level and their support. Further, in the standard models under iid Gaussian errors, the Bayesian shrinkage rules typically outperform the thresholding rules in terms of averaged mean square errors, see \cite{angelini-vidakovic-2004}, \cite{sousa-2020} and \cite{sousa-et-al-2020}.

\cite{sousa-2025} proposed a Bayesian wavelet shrinkage rule based on a prior for the wavelet coefficients which is composed of a mixture of a point mass function at zero and the logistic distribution. Their simulation studies indicated, overall, strong performance of this wavelet shrinkage rule compared to the universal thresholding rule proposed by \cite{johnstone-silerman-1997} in models with autoregressive AR(1) and ARFIMA errors. Despite its effectiveness, an additional step in the wavelet shrinkage process, involving differencing the original data before applying the shrinkage/thresholding rule, can applied to reduce the correlation of the errors and improve estimation of the coefficients.

Thus, compared to \cite{sousa-2025}, we propose a different approach to wavelet shrinkage, applying it to the differenced data rather than directly to the original data. Specifically, we apply the Bayesian shrinkage rule proposed by \cite{sousa-2025} to the differenced data and then we recover the estimated signal. This approach minimizes the effect of the strong-dependence (long-memory) of the errors on the estimation. 

The remainder of this paper is organized as follows: the statistical model with ARFIMA errors and the proposed wavelet shrinkage procedure are described in Section \ref{sec:method}. Simulation studies evaluating the performance of the proposed method are discussed in Section \ref{sec:simulations}. An application to a real dataset involving temperature anomalies in the Northern Hemisphere is presented in Section \ref{sec:application}. Final considerations are provided in Section \ref{sec:conclusions}.

\section{Statistical model and proposed method} \label{sec:method}
We consider $n+1$ observations $(x_1,y_1),\ldots,(x_{n+1},y_{n+1})$, $n = 2^J$, $J \in \mathbb{N}$, from the unidimensional nonparametric regression model
\begin{equation}\label{timemodel}
y_i = f(x_i) + e_i,
\end{equation}
where $x_1,\ldots,x_{n+1}$ are equally spaced scalars and $f$ is an unknown squared integrable function, i.e, $f \in \mathbb{L}^2(\mathbb{R})$. Also, we assume that $e_1,\ldots,e_{n+1}$ are random errors following an ARFIMA(0,$d$,0) process:
\begin{equation}\label{arfima-errors}
(1-B)^d e_i = a_i
\end{equation}
where $B$ is backshift operator, $(1-B)^d=\sum_{j=0}^\infty b_j(d)B^j$ with $b_j(d) = \Gamma(j-d)/(\Gamma(j+1)\Gamma(-d))$ for $j=0,1,\ldots$, with $\Gamma(\cdot)$ being the gamma function, and 
$\{a_i\}$ is an iid sequence with zero mean and finite variance $\sigma^2_a$. Additionally, to reproduce very long-memory in the error sequence, we assume that the parameter $d$ satisfies $0<d<0.5$.

The goal is to estimate $f$. To do this, the standard wavelet-based procedure is to apply a discrete wavelet transform (DWT) on the original vector of observations to obtain the empirical wavelet coefficients. Then, a thresholding or shrinkage rule is applied on the empirical wavelet coefficients in order to estimate the wavelet coefficients. Based on these coefficients, the function values are then estimated by applying the inverse discrete wavelet transform (IDWT). See \cite{vidakovic-1999} and \cite{nason-2008} for more details.

Next, we describe our approach. In this paper we propose the application of the DWT to the vector of differences of the observations, i.e
\begin{equation}\label{dif}
    z_i = y_{i+1} - y_i = (1-B)y_i, \hspace{0.5cm} i=1,\ldots,n.
\end{equation}
Then, from \eqref{timemodel} and \eqref{dif}, we have  
\begin{equation}\label{difmodel}
    z_i = g(x_i,x_{i+1}) + \eta_i, \hspace{0.5cm} i=1,\ldots,n,
\end{equation}
where $g(x_i,x_{i+1}) = f(x_{i+1}) - f(x_i)$ is the signal of the observations $z_i$ and $\{\eta_i\}$ with $\eta_i = e_{i+1} - e_{i}$ is a new error sequence which follows an ARFIMA(0,$d-1$,0) process\footnote{Note that $\eta_i = (1-B)e_{i+1}$, and from \eqref{arfima-errors} $e_{i+1} = (1-B)^{-d}a_{i+1}$, then $\eta_i = (1-B)(1-B)^{-d}a_{i+1}$. Therefore, $(1-B)^{d-1}\eta_i= a_{i+1}$.}. Since $0<d<0.5$, the sequence $\{\eta_i\}$ exhibits much less memory compared to the original error sequence $\{e_i\}$.

Now, we can rewrite model \eqref{difmodel} in vector notation as
\begin{equation}\label{difmodelvec}
    \boldsymbol{z} = \boldsymbol{g} + \boldsymbol{\eta},
\end{equation}
where $\boldsymbol{z} = [z_1,\cdots,z_n]'$, $\boldsymbol{g} = [g(x_1,x_2),\cdots,g(x_n,x_{n+1})]'$ and $\boldsymbol{\eta} = [\eta_1,\cdots,\eta_n]'$. The DWT can be represented by an $n \times n$ orthogonal matrix $W$ that is applied to both sides of equation \eqref{difmodelvec} although it is applied in practice by the pyramidal algorithm, obtaining the model in the wavelet domain
\begin{equation}\label{wavemodel}
    \boldsymbol{w} = \boldsymbol{\theta} + \boldsymbol{\varepsilon},
\end{equation}
where $\boldsymbol{w}= W\boldsymbol{z} = [w_1,\cdots,w_n]'$ is the vector of empirical wavelet coefficients, $\boldsymbol{\theta} = W\boldsymbol{g}=[\theta_1,\cdots,\theta_n]'$ is the vector of the wavelet coefficients and $\boldsymbol{\varepsilon} = W\boldsymbol{\eta}= [\varepsilon_1,\cdots,\varepsilon_n]'$ is the vector of random errors in the wavelet domain. Then, the single wavelet coefficient $\theta$ is estimated by applying a thresholding or shrinkage rule $\delta(\cdot)$ to its associated empirical coefficient $w$, i.e, 
\begin{equation}\label{shrink}
    \hat{\theta} = \delta(w).
\end{equation}
Finally, the function $g$ is estimated by applying the inverse DWT (IDWT) to the estimated vector of wavelet coefficients
\begin{equation}\label{estg}
    \boldsymbol{\hat{g}} = W^{'}\boldsymbol{\hat{\theta}},
\end{equation}
and the function $f$ is then estimated by
\begin{equation}\label{estf}
    \hat{f}(x_{i+1}) = \hat{f}(x_{i}) + \hat{g}(x_i,x_{i+1}), \hspace{0.5cm} i=1,\ldots,n,
\end{equation}
where $\hat{f}(x_1)$ has to be estimated. In this paper, we use a simple estimator: $\hat{f}(x_1)=(1/20)\sum_{i=1}^{20} y_i$, i.e, the mean of the first 20 observations. 

Furthermore, we consider the resolution level dependent Bayesian shrinkage rule proposed by \cite{sousa-2025} which estimates a given wavelet coefficient $\theta$ of the resolution level $j$, $j = 0,\ldots, J-1$, by 
\begin{equation}\label{rule}
\hat{\theta} = \delta_j(w)= \frac{(1-\alpha)\int_\mathbb{R}(\hat{\sigma}_j u + w)h(\hat{\sigma}_j u +w ; \tau)\phi(u)du}{\frac{\alpha}{\hat{\sigma}_j}\phi(\frac{w}{\hat{\sigma}_j})+(1-\alpha)\int_\mathbb{R}h(\hat{\sigma}_j u +w ; \tau)\phi(u)du},
\end{equation}
where $\alpha \in (0,1)$, $\phi(\cdot)$ is the standard normal density function, $\hat{\sigma}_j$ is an estimate of the standard deviation of the errors at resolution level $j$ and $h(\cdot;\tau)$ is the logistic density function symmetric around zero
\begin{equation}
h(\theta;\tau) = \frac{\exp\{-\theta/\tau\}}{\tau(1 + \exp\{-\theta/\tau\})^2}\mathbb{I}_{\mathbb{R}}(\theta), \nonumber
\end{equation}
for $\tau > 0$. The choice of this wavelet estimator was motivated by the good results obtained in Monte Carlo experiments when the Bayesian shrinkage rule \eqref{rule} was compared to the soft thresholding rule proposed by \cite{johnstone-silerman-1997} when errors $e_i$ follow AR(1) and ARFIMA processes; see \cite{sousa-2025}.

\begin{figure}
    \centering
    \includegraphics[width=1\linewidth]{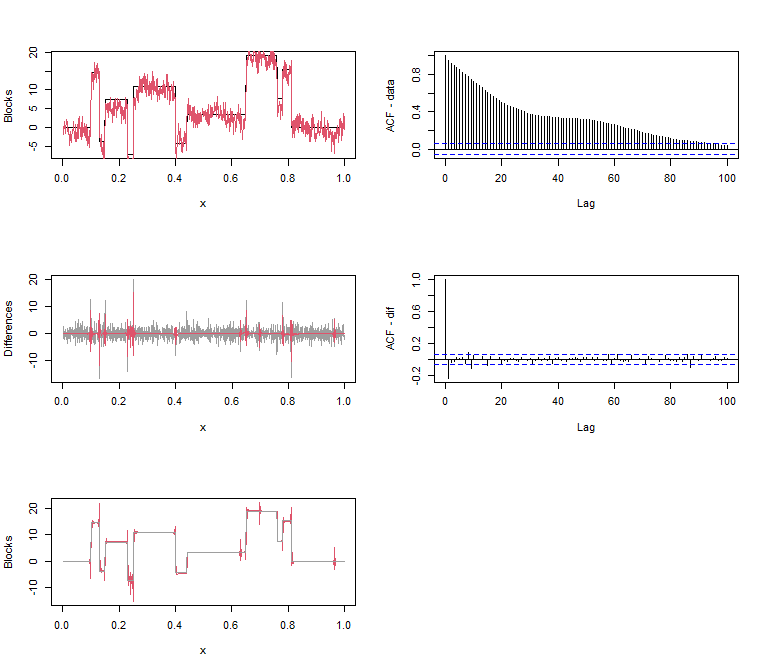}
    \caption{Underlying Blocks function and generated data at the top-left; sample autocorrelation function of the generated data at the top-right; differenced data and their denoised versions by  application of the shrinkage rule at the center-left; sample autocorrelation function of the differenced data at the center-right; and estimated signal function by the proposed method at the bottom-left.}
    \label{fig:example}
\end{figure}

We illustrate the proposed method with an example that involves the Donoho-Johnstone test function called Blocks, which is represented in the top left of Figure \ref{fig:example} in the black line. We generated $1025 = 2^{10} + 1$ observations from the Blocks function ($f$) and added random errors following an ARFIMA(0,$d$,0) process with $d=0.4$ via Equation \eqref{timemodel}. The variance of the errors, $\sigma^2_a$, was chosen according to a signal-to-noise ratio (SNR) equal to 3. The generated dataset is represented in the top-left of Figure \ref{fig:example} in the red line and its sample autocorrelation function (ACF) is shown in the top-right of Figure \ref{fig:example}. Note that the ACF clearly exhibits a long-memory behavior.

\vspace{0.3cm}
\centerline{[Figure \ref{fig:example} around here]}
\vspace{0.3cm}

Figure \ref{fig:example} shows in the center-left the differenced data ($z_i$) in grey, and the corresponding shrunk version in red, obtained by applying the logistic shrinkage rule \eqref{rule} to the wavelet domain of the empirical wavelet coefficients of the differences. Additionally, the center-right of Figure \ref{fig:example} presents the ACF of the differences $z_i$. In this ACF there is evident reduction of the memory compared to the ACF of the original dataset. Finally the graph on the bottom-left of Figure \ref{fig:example} represents the estimated signal in the red line. Note that the estimated signal function recovers well the main features of the Blocks function such its discontinuities and constant parts, even when we have low SNR in the data.

\section{Simulation studies} \label{sec:simulations}

We conducted simulation studies to evaluate the performance of the proposed method (Log-Diff) against the universal thresholding rule of \cite{johnstone-silerman-1997} (Universal) and the Bayesian shrinkage rule under a logistic prior of \cite{sousa-2025} (Logistic), all designed to handle models with correlated noise. The so-called Donoho-Johnstone test functions: Bumps, Blocks, Doppler, and Heavisine, as proposed by \cite{donoho-johnstone-1994}, were considered as the underlying functions to be estimated.

For each underlying function, we generated data from model \eqref{timemodel}-\eqref{arfima-errors} with $d=0.4$ using two sample sizes: $n = 512$ and $n = 2048$, and two signal-to-noise ratios: $\mathrm{SNR} = 3$ and $\mathrm{SNR} = 9$. After generating the data, the signal function was estimated using the three methods, and the mean squared error (MSE) was calculated for each method. The process was replicated 200 times, and the averaged mean squared error (AMSE) was then computed for each method as follows: 
\begin{equation} \mathrm{AMSE} = \frac{1}{200n}\sum_{r = 1}^{200}\sum_{i = 1}^{n}[\hat{f}_r(x_i) - f(x_i)]^2, \nonumber \end{equation} 
where $\hat{f}_r(x_i)$ is the estimate of $f(x_i)$ in replication $r$. The standard deviation of the MSEs was also calculated and is an important aspect of the analysis.

Table \ref{tab:sim} presents the results. Here, we observe that according to the AMSE, the Bayesian logistic shrinkage rule and the proposed method based on differences performed best, except for the Heavisine signal when SNR=3, in which case the Universal method was the best. In fact, the Log-Diff and Logistics methods exhibited similar performances across all scenarios. However, the Log-Diff method prevailed over Bumps and Blocks, except when SNR=3 and $n=2048$. Besides this, for the Doppler function, the Log-Diff method is the best or second best. 

\vspace{0.3cm}
\centerline{[Table \ref{tab:sim} around here]}
\vspace{0.3cm}

\begin{figure}
    \centering
    \includegraphics[width=1\linewidth]{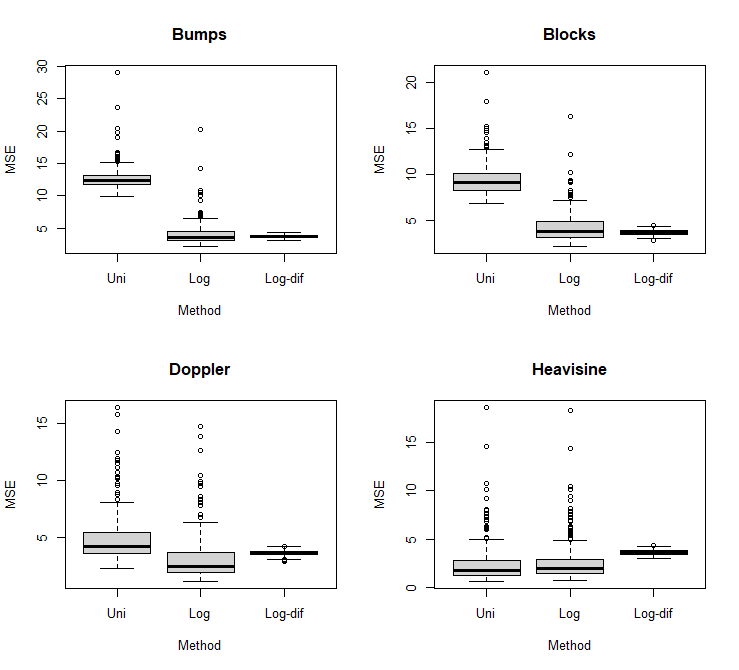}
    \caption{Boxplots of the MSEs of estimations by the universal thresholding (Uni), Bayesian logistic shrinkage rule (Log) and Bayesian logistic shrinkage in the differences (Log-dif) for the Donoho-Johnstone test functions with $n = 512$ and $\mathrm{SNR = 3}$.}
    \label{fig:bp}
\end{figure}

Although the proposed method and the Bayesian logistic rule were close in terms of AMSE, they differed in the standard deviations of the MSEs. In fact, the MSEs of the proposed method exhibited smaller dispersion than those of the logistic rule, and much smaller dispersion than the Universal rule, in all scenarios. For instance, considering the Bumps function with $n = 2048$, the two methods were similar in terms of AMSE, as mentioned earlier, but the standard deviation of the proposed method was 0.122 compared to 1.493 for the logistic rule when $\mathrm{SNR} = 3$—about 12.24 times larger. When $\mathrm{SNR} = 9$, the standard deviation was 0.014 for the proposed method compared to 0.155 for the logistic rule—about 11.07 times larger. This advantage of the proposed method is also evident in the boxplots of the MSEs for $n = 512$ and $\mathrm{SNR} = 3$, provided in Figure \ref{fig:bp}. These boxplots clearly show the better precision of the estimates from the proposed method compared to both the universal thresholding rule and the Bayesian logistic rule. Similar behavior was observed in the boxplots constructed for the other scenarios of sample size and SNR.

\vspace{0.3cm}
\centerline{[Figure \ref{fig:bp} around here]}
\vspace{0.3cm}

\section{Real data illustration} \label{sec:application}

We applied the proposed method to smooth a time series involving the monthly Northern Hemisphere temperature anomalies (in ºC) from January 1880 to May 1965 ($n + 1 = 1025$). This dataset, available in the R package \textit{smoots} of \cite{feng-2023}, was analyzed by \cite{feng-2007} and \cite{sousa-2025} in the context of nonparametric regression with ARFIMA(0,$d$,0) errors. 

\begin{figure}
    \centering
    \includegraphics[width=1\linewidth]{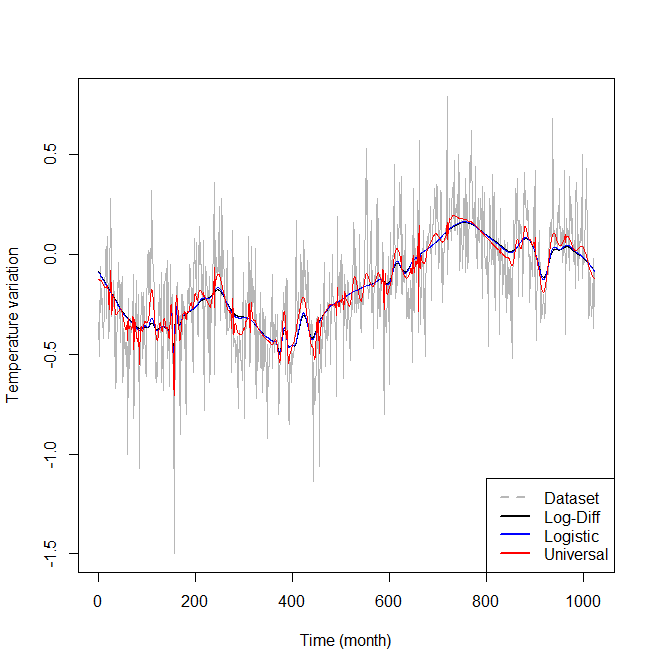}
    \caption{Monthly Northern Hemisphere temperature anomalies (in ºC) from January 1880 to May 1965 and the smoothed versions obtained by the proposed method (Log-Diff), Logistic and Universal methods.}
    \label{fig:app}
\end{figure}

Figure \ref{fig:app} shows the dataset and the estimated curves obtained using the proposed method (Log-Diff) and the Logistic and Universal methods. Note that the estimated functions successfully capture the main characteristics of the data, but the Log-Diff and Logistic methods (which are very close) provide significant noise reduction. The estimated SNR is 0.89, computed by approximating the standard deviation of the signal (the unknown function $f$) using the estimated curve $\hat{f}$ obtained with the proposed method, indicating a substantial level of noise in the data.  

\vspace{0.3cm}
\centerline{[Figure \ref{fig:app} around here]}
\vspace{0.3cm}

\section{Conclusions} \label{sec:conclusions}

In this paper, we propose a method for curve estimation, based on wavelet Bayesian shrinkage, in nonparametric regression models with long-memory correlated errors. This method modifies the proposal of \cite{sousa-2025}, using transformed (differenced) data to estimate the signal function. The differentiation of the original data permits obtaining a new model with noise that has much lower correlation compared to the original noise. 

Simulation studies showed that in terms of averaged mean squared error, the proposed procedure had similar behavior compared to the Bayesian shrinkage rule under logistic prior applied to empirical coefficients of the data (without differencing) but was better than the Johnstone and Silverman thresholding rule. However, in terms of precision, measured by the standard deviation of the mean squared errors, the proposal is better than the mentioned methods in all scenarios.


\bibliographystyle{plainnat}
\bibliography{references}

\begin{thebibliography}{16}
\providecommand{\natexlab}[1]{#1}
\providecommand{\url}[1]{\texttt{#1}}
\expandafter\ifx\csname urlstyle\endcsname\relax
  \providecommand{\doi}[1]{doi: #1}\else
  \providecommand{\doi}{doi: \begingroup \urlstyle{rm}\Url}\fi

\bibitem[Angelini and Vidakovic(2004)]{angelini-vidakovic-2004}
C.~Angelini and B.~Vidakovic.
\newblock Gama-minimax wavelet shrinkage: a robust incorporation of information about energy of a signal in denoising applications.
\newblock \emph{Statistica Sinica}, 14\penalty0 (1):\penalty0 103--125, 2004.

\bibitem[Deo(1997)]{deo-1997}
R.S. Deo.
\newblock Nonparametric regression with long-memory errors.
\newblock \emph{Statistics and Probability Letters}, 33\penalty0 (1):\penalty0 89--94, 1997.

\bibitem[Donoho and Johnstone(1994)]{donoho-johnstone-1994}
D.L. Donoho and I.M. Johnstone.
\newblock Ideal spatial adaptation by wavelet shrinkage.
\newblock \emph{Biometrika}, 81\penalty0 (1):\penalty0 425--455, 1994.

\bibitem[Feng(2007)]{feng-2007}
Y.~Feng.
\newblock On the asymptotic variance in nonparametric regression with fractional time-series errors.
\newblock \emph{Journal of Nonparametric Statistics}, 19\penalty0 (2):\penalty0 63--76, 2007.

\bibitem[Feng et~al.(2023)Feng, Letmathe, and Schulz]{feng-2023}
Yuanhua Feng, Sebastian Letmathe, and Dominik Schulz.
\newblock \emph{smoots: Nonparametric Estimation of the Trend and Its Derivatives in TS}, 2023.
\newblock URL \url{https://CRAN.R-project.org/package=smoots}.
\newblock R package version 1.1.4.

\bibitem[Granger and Joyeux(1980)]{granger-joyeux-1980}
C.~W.~J. Granger and R.~Joyeux.
\newblock An introduction to long‐memory time series models and fractional differencing.
\newblock \emph{Journal of Time Series Analysis}, 1\penalty0 (1):\penalty0 15--29, 1980.

\bibitem[Hall and Hart(1990)]{hall-1990}
P.~Hall and J.D. Hart.
\newblock Nonparametric regression with long-range dependence.
\newblock \emph{Stochastic Processes and their Applications}, 36\penalty0 (1):\penalty0 339--351, 1990.

\bibitem[Hosking(1981)]{hosking-1981}
J.R.M. Hosking.
\newblock Fractional differencing.
\newblock \emph{Biometrika}, 68\penalty0 (1):\penalty0 165--176, 1981.

\bibitem[Johnstone and Silverman(1997)]{johnstone-silerman-1997}
I.M. Johnstone and B.W. Silverman.
\newblock Wavelet threshold estimators for data with correlated noise.
\newblock \emph{Journal of the Royal Statistical Society B}, 59\penalty0 (2):\penalty0 319--351, 1997.

\bibitem[Nason(2008)]{nason-2008}
G.~P. Nason.
\newblock \emph{Wavelet Methods in Statistics with R}.
\newblock Springer, New York, 2008.

\bibitem[Porto et~al.(2016)Porto, Morettin, Percival, and Aubin]{porto-et-al-2016}
R.~Porto, P.A. Morettin, D.~Percival, and E.~Aubin.
\newblock Wavelet shrinkage for regression models with random design and correlated errors.
\newblock \emph{Brazilian Journal of Probability and Statistics}, 30\penalty0 (4):\penalty0 614--652, 2016.

\bibitem[Sousa(2020)]{sousa-2020}
A.R.S. Sousa.
\newblock Bayesian wavelet shrinkage with logistic prior.
\newblock \emph{Communications in Statistics - Simulation and Computation}, 51\penalty0 (8):\penalty0 4700--4714, 2020.

\bibitem[Sousa and Zevallos(2025)]{sousa-2025}
A.R.S. Sousa and M.~Zevallos.
\newblock On bayesian wavelet shrinkage estimation of nonparametric regression models with stationary correlated noise.
\newblock \emph{Statistics and Computing}, 35\penalty0 (83), 2025.

\bibitem[Sousa et~al.(2020)Sousa, Garcia, and Vidakovic]{sousa-et-al-2020}
A.R.S. Sousa, N.L. Garcia, and B.~Vidakovic.
\newblock Bayesian wavelet shrinkage with beta prior.
\newblock \emph{Computational Statistics}, 36\penalty0 (2):\penalty0 1341--1363, 2020.

\bibitem[Vidakovic(1999)]{vidakovic-1999}
B.~Vidakovic.
\newblock \emph{Statistical Modeling by Wavelets}.
\newblock Wiley, New York, 1999.

\bibitem[Wang(1996)]{wang-1996}
Y.~Wang.
\newblock Function estimation via wavelet shrinkage for long-memory data.
\newblock \emph{The Annals of Statistics}, 24\penalty0 (2):\penalty0 466--484, 1996.

\end{thebibliography}

\clearpage

\begin{table}[H]
\centering
\label{my-label}
\begin{tabular}{crlcc}
\hline
Signal & $n$ & Method & SNR = 3 & SNR = 9 \\ 
&& & AMSE (SD)& AMSE (SD) \\ \hline \hline
Bumps & 512 & Universal & 12.921 (2.150) & 2.247 (0.248) \\
      &     & Logistic & 4.147 (1.975) & 0.623 (0.228) \\
      &     & Log-Diff & \textbf{3.666 (0.237)} &  \textbf{0.399 (0.027)} \\ 
     & 2048 & Universal & 6.929 (1.598) & 1.140 (0.156) \\
     &      & Logistic & \textbf{3.312} (1.493) & 0.413 (0.155) \\
     &      & Log-Diff & 3.642 (\textbf{0.122}) & \textbf{0.401 (0.014)} \\ 
\\[-0.5em]
Blocks & 512 & Universal & 9.454 (1.918) & 1.855 (0.256) \\
      &      & Logistic & 4.320 (1.887) & 0.582 (0.243) \\
      &      & Log-Diff & \textbf{3.618 (0.268)} & \textbf{0.403 (0.027)} \\ 

     & 2048 & Universal & 5.367 (1.564) & 1.140 (0.230) \\
     &      & Logistic & \textbf{3.221} (1.528) & 0.442 (0.220) \\
     &     & Log-Diff & 3.629 (\textbf{0.132}) & \textbf{0.402 (0.015)} \\ 
\\[-0.5em]
Doppler & 512 & Universal & 4.998 (2.348) & 0.846 (0.202) \\
       &      & Logistic & \textbf{3.285} (2.281) & 0.408 (0.199) \\
       &      & Log-Diff &3.627 (\textbf{0.239}) & \textbf{0.407 (0.031)} \\ 

       & 2048 & Universal & 2.831 (1.468) & 0.478 (0.222) \\
       &      & Logistic & \textbf{2.627} (1.436) & \textbf{0.332} (0.221) \\
       &      & Log-Diff & 3.629 (\textbf{0.116}) & 0.401 (\textbf{0.014}) \\ 
\\[-0.5em]
Heavisine & 512 & Universal & \textbf{2.597} (2.383) & 0.463 (0.222) \\
         &      & Logistic & 2.767 (2.299) & \textbf{0.360} (0.219) \\
         &      & Log-Diff & 3.650 (\textbf{0.263}) & 0.413 (\textbf{0.028}) \\ 

         & 2048 & Universal & \textbf{1.922} (1.620) & 0.333 (0.194) \\
         &      & Logistic & 2.474 (1.613) & \textbf{0.313} (0.194) \\
         &      & Log-Diff & 3.634 (\textbf{0.121}) & 0.401 (\textbf{0.014}) \\ \hline
\end{tabular}
\caption{AMSE and standard deviation of MSE (SD) of the estimations by the universal thresholding (Universal), Bayesian logistic shrinkage rule (Logistic) and the proposed Bayesian logistic shrinkage in the differences (Log-Diff) for the Donoho-Johnstone test functions. For each combination of signal, $n$ and SNR, the best results of AMSE and SD are marked in bold.} \label{tab:sim}
\end{table}


\clearpage

\end{document}